\documentclass[9pt,twocolumn,twoside]{osajnl}
\journal{ao} % Choose journal (ao,jocn,josaa,josab,ol,optica,pr)

\setboolean{shortarticle}{false}
\usepackage{subfig}
\usepackage{siunitx}
\usepackage[utf8]{inputenc}
\usepackage{amsmath,amsfonts,amssymb}
\title{Method for tilt correction of calibration lines in high resolution spectra}

\author[1,2,*]{Tanya Das}
\author[1]{Ravinder K. Banyal}

\affil[1]{Indian Institute of Astrophysics, Bangalore, India}
\affil[2]{University of Calcutta, JD-2 Salt Lake, Kolkata 750098, India}

\affil[*]{Corresponding author: tanya@iiap.res.in}

\begin{abstract}
Technological advancement has led to improvement in the design capabilities of astronomical spectrographs, allowing for high precision spectroscopy, thereby expanding the realms of observational astronomy. High-resolution spectrographs use Echelle grating that operates in higher orders, giving more detailed spectra. Often, curvature and tilted lines are observed in the spectra, arising due to the design trade-offs of the respective spectrographs. Removal of these artifacts can help avoid wrong flux calculation and line centroid position misinterpretation, which can aid in a better prediction of the wavelength calibration model. In this paper we present a post-processing technique that we developed to correct the observed curvature and tilt in the spectra. We have demonstrated the correction technique on Fabry-Perot and Th-Ar calibration spectra obtained from Hanle Echelle Spectrograph (HESP), Magellan Inamori Kyocera Echelle (MIKE) spectrometer and X-shooter spectrograph.
\end{abstract}

\setboolean{displaycopyright}{true}

\begin{document}

\maketitle

\section{Introduction}
\label{sec:intro}
Spectroscopy is a key method for monitoring and analyzing the physical and chemical properties of astronomical objects. Measurement of the dark matter content of galaxies and their clusters \cite{ref:darkmatter}, estimation of the mass of stellar systems \cite{ref:massratio}, determination of the age of stars \cite{ref:age}, identification of chemical composition, temperature and other parameters by studying the strength of spectral features \cite{ref:anirban} and radial velocity measurements to determine the presence of planets around stars \cite{ref:hatzes} are the primary fields of application for spectroscopy.

Spectrographs are instruments used to observe the spectrum of astrophysical objects.  Usually, conventional low-resolution  spectrographs are compact and light weight. They are directly attached to Cassegrain focus of the telescope behind the primary mirror. The direct interface with telescope makes it easier to focus the star light directly onto the spectrograph slit. However, high-resolution Echelle spectrographs (typically $R \gtrsim 50,000$) tend to be bulky and require larger space. In such cases, spectrograph is decoupled from the main telescope and  placed on a stable Nasmyth platform or sometimes housed in a separate room. A tertiary mirror can be used to steer the telescope beam onto the Nasmyth focus. Alternatively, fibers are deployed to collect the light from telescope focal plane and deliver it to the spectrograph slit. A diverging beam from the slit is collimated and then passed onto the dispersing element.  A diffraction grating is used in most spectrographs to disperse the light from the target object into its component wavelengths. A collimated beam incident on the grating ensures that different wavelengths are dispersed at distinct angles. The camera optics then focuses the dispersed light on to a charged coupled device (CCD), where it is recorded for further analysis.

In contrast to their low-resolution counterparts, high-resolution spectrographs employ Echelle grating for use in higher diffraction order, thereby providing more details about spectral features and giving resolution up to R = 150,000. Echelle gratings have large blaze angles, typically between 50$^{\circ}$-75$^{\circ}$, providing dispersion at higher angles, and optimized to concentrate maximum efficiency in a specific direction. At higher orders, they offer substantial overlap, causing the longer wavelength of a higher order to overlap with the following order's shorter wavelength. A cross disperser, like a prism or a grating, is used to separate the overlapping orders in spatial direction to achieve a wider spectral coverage.

Stellar spectra consist of both absorption and emission lines, whose precise position is determined by converting the pixels into the wavelength scale. This step is described as wavelength calibration, which utilizes known laboratory sources such as Th-Ar or advanced techniques such as Fabry-Perot (FP) etalon \cite{ref:wildi}, Iodine cell \cite{ref:marcy}, and laser frequency comb \cite{ref:murphy}.

The benefits provided by echelle spectrographs also come with some shortcomings and design constraints. In order to accommodate the spectra effectively on the detector, optical components, such as slit or camera, are often adjusted, leading to the introduction of artifacts and aberration like distortion, defocus and tilted lines. The grating also introduces an out-of-plane gamma angle, along with cross disperser prism, which induces tilt in the spectral lines \cite{ref:siri_stb}. Neglecting these factors may lead to addition of errors.

\begin{figure}[htbp]
\centering
\vbox{\includegraphics[width=0.9\linewidth]{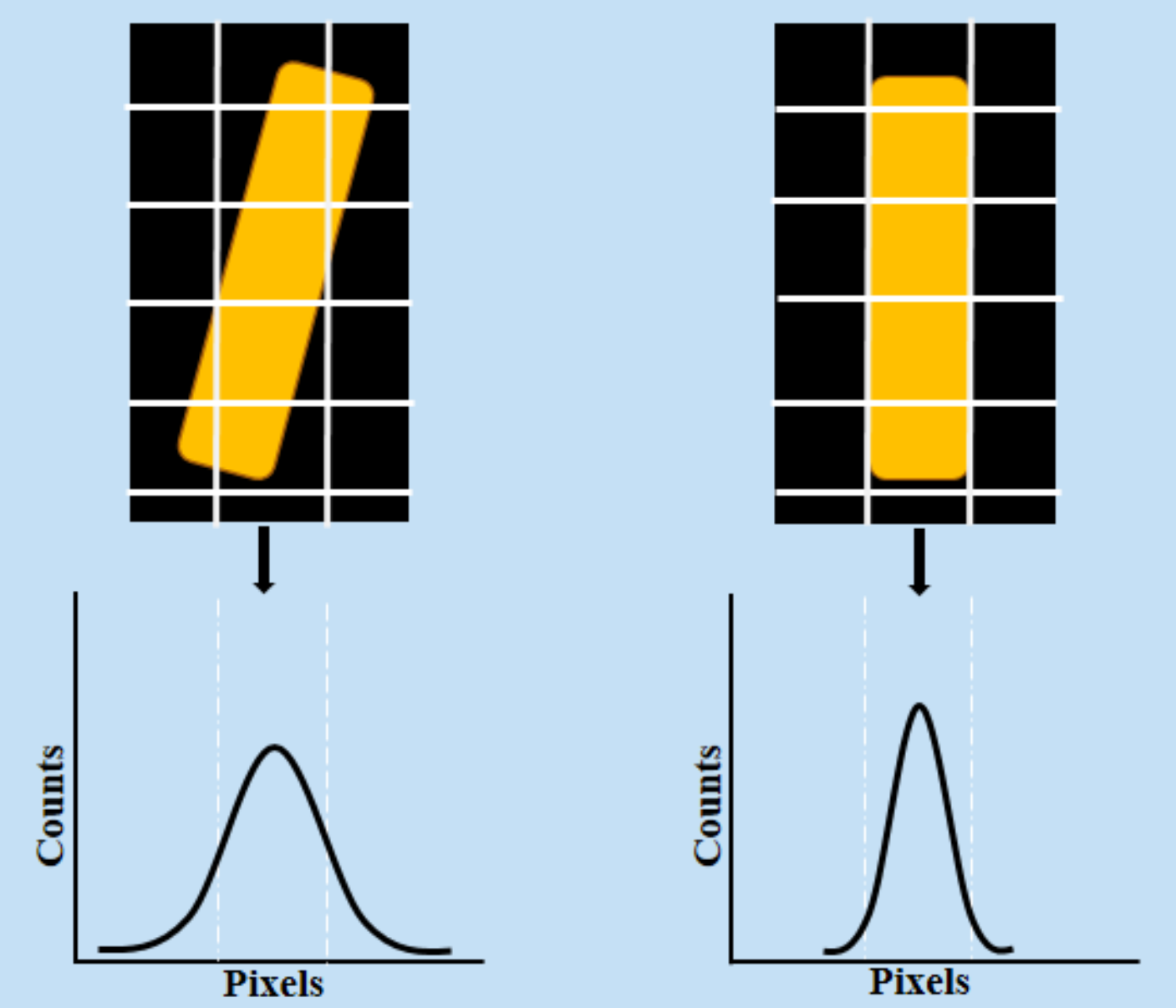}}
\caption{A cartoon depicting the effect of tilt on the FWHM of the lines.}
\label{fig:tiltshow}
\end{figure}

The tilt in calibration and spectral lines is also introduced by imperfections resulting from various trade offs in the spectrograph design. Curvature in the spectrograph orders can arise due to the cross disperser, whose dispersion direction is perpendicular to the Echelle grating. The tilt in individual lines in the spectrum are caused due to the cross disperser and Echelle grating working on two different planes (operating in quasi-Littrow mode). Generally, the binning of data is done normal to the dispersion axis, and hence if the tilt in spectral lines is not considered, the resulting 1D data may show an increase in the FWHM of the lines and also cause blending in some cases. This is because, in the case of tilted lines, the intensity is distributed over several pixels, which, when binned along the slit direction, gives a broadened line and wrong flux values, as shown in figure \ref{fig:tiltshow}. The broadened spectral line also results in the degradation of spectral resolution. The tilt in the slit image at the detector is a function of wavelength \cite{ref:tiltwav}. It also compromises the attainable accuracy in RV measurements if not adequately taken care of or modelled out by the extraction software \cite{ref:ireland}. Disentangling the tilt related shifts in line centroid position from the RV shift is essential for RV studies. The point spread function (PSF) is predominantly the response of the instrument to a monochromatic light or a point source. The instrumental aberrations of the system already broaden the energy distribution in the PSF. The tilt is an artifact of the system, which causes further broadening of the line. Removal of this artifact can hence facilitate the effort of predicting instrument PSF by removing one of the dependent factors, in cases where it is necessary to generate a PSF map of the instrument and its variation during an observation run. Measuring the tilt can also help in the study and estimation of instrument aberrations, for example, a better prediction of distortion in the system.

In this paper, we are using FP and Th-Ar calibration data to assess and remove the tilt and curvature being introduced by the instrument. Since FP is a stable source that provides several continuous lines of equal intensity, it helps in better estimation and elimination of these artifacts, thereby resulting in a better prediction of the wavelength calibration model. We have designed and developed an FP based wavelength calibration system \cite{ref:fpspie}. The system was tested on Hanle Echelle Spectrograph (HESP), where the artifacts of the instrument often show up in the FP spectra. Spectrographs like Magellan Inamori Kyocera Echelle (MIKE) spectrometer \cite{ref:mike} and X-shooter \cite{ref:xshooter} show highly tilted spectral lines due to the quasi-Littrow configuration. The tilt removal algorithm developed as part of this study has been tested on X-Shooter and MIKE Th-Ar calibration data. A general review of the existing methods adopted for curvature and tilt correction in Echelle spectra is presented in \ref{sec:existing}. The proposed methodology and algorithm are described in section \ref{sec:method}. Results from the correction algorithm have been presented in section \ref{sec:result} followed by a summary of this work in section \ref{sec:summary}.

\section{Review of existing techniques for tilt and curvature removal}
\label{sec:existing}

In order to obtain calibrated science quality data, the two-dimensional spectrograms obtained from the spectrograph have to be processed and the spectrum extracted. Software like IRAF \cite{ref:iraf86, ref:iraf93} can be used for general reduction purpose. Many spectrographs have dedicated state of the art pipeline for preparation and reduction of the 2D spectra. A brief description of the method adopted in X-Shooter and MIKE pipeline will be discussed in this section, along with a new algorithm called PyReduce developed for the same purpose.

\begin{figure}[htbp]
\centering
\vbox{\includegraphics[width=0.98\linewidth]{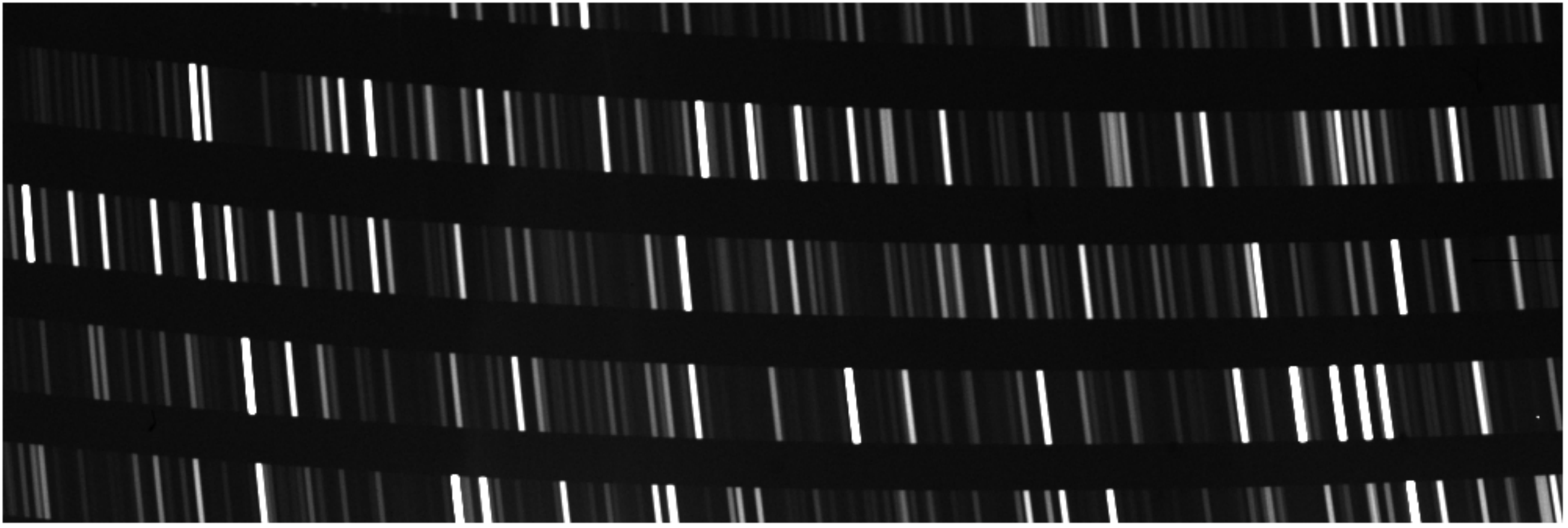}}
\caption{Zoomed in Th-Ar calibration spectra from X-shooter showing highly curved orders along with large tilt in spectral lines.}
\label{fig:xshspectra}
\end{figure}

X-Shooter is a single target spectrograph of medium resolution (R$\sim$ 4000-17000), installed at the Cassegrain focus of ESO's Very Large Telescope (VLT) \cite{ref:vernet}. The output of this spectrograph consists of highly curved orders with tilted spectral lines, shown in figure \ref{fig:xshspectra}. Due to this, special measures are taken for wavelength calibration and optimal extraction of the spectra, as described in \cite{ref:goldoni06, ref:goldoni08}. X-Shooter pipeline is written in ANSI C. Separate pinhole mask arc frames are taken for computation of initial guess for wavelength solution and determine spatial and wavelength scale calibrations. A polynomial interpolation, constructed using multi-pinhole frame, is used for transforming detector coordinates into a function of wavelength (determined by the guess solution), order number and position on the slit that allows the removal of order curvature and line tilts. The physical model of the instrument can also be used to determine the wavelength position on detector. Finally, the detector pixels are oversampled in 2D and linear interpolation is used to find the slit profile which is then collapsed over user defined slits. A detailed description of the entire pipeline can be found in \cite{ref:modigliani}.

\begin{figure}[htbp]
\centering
\vbox{\includegraphics[width=\linewidth]{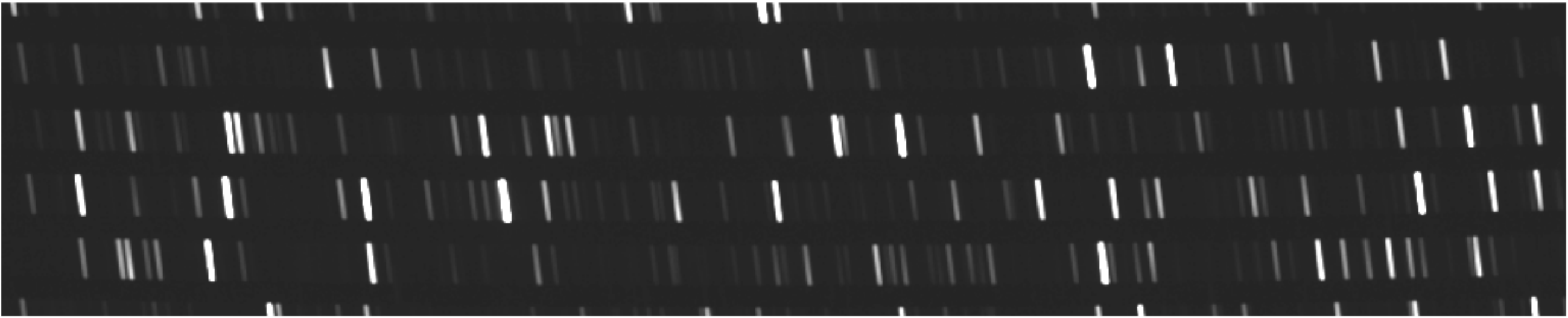}}
\caption{Zoomed in Th-Ar calibration spectra from red channel of MIKE at 0.7" slit setting, showing tilted lines.}
\label{fig:mikespectra}
\end{figure}

The Magellan Inamori Kyocera Echelle (MIKE) is a high resolution, double echelle spectrograph installed on the Magellan II telescope at Las Campanas Observatory, Chile. Figure \ref{fig:mikespectra} shows the Th-Ar calibration spectra as captured in the red arm for a slit size of 0.7 inches. A dedicated pipeline is available for data reduction \cite{ref:reduc} using IDL with a Python alternative. After the standard image processing involving overscan removal and flat fielding, a 2D wavelength image is generated by using Th-Ar arc frame to derive a 1D wavelength solution along the center of each order. For each order, the high SNR arc lines are identified and their tilt is measured as a function of wavelength and echelle order. The centroid of each line is traced across the order and best fit line for each arc line is used to calculate the slope. A 2D Legendre polynomial is fit to the calculated slopes for every order and the slope values interpolated from the solution for areas with less density of Th-Ar lines. A unique wavelength is assigned to the center of all the pixels falling within the echelle order by using the derived wavelength solution and the arc line tilts. This is done across full CCD pixels and a wavelength image is generated which is finally used in the optimal extraction of the orders. A similar method is adopted by MAGE Spectral Extractor (MASE) \cite{ref:bochanski}.

The new REDUCE package \cite{ref:pyreduce} adds to the earlier version of the developed package \cite{ref:valenti} by incorporating tilted and curved slit images. The method follows a slit decomposition algorithm, where the 2D image of the spectral order is represented by slit illumination and spectrum, sampled on the detector. The shape of slit image is modelled by taking strong and unblended emission lines in a wavelength calibrated spectrum and fitting a 2D Gaussian to each of the line image. The tilt and curvature variations across the order is combined by fitting a polynomial and interpolating to all the columns. Optimal extraction is then performed keeping in mind the calculated tilt and curvature.

\begin{figure}[htbp]
\centering
\fbox{\includegraphics[width=0.8\linewidth]{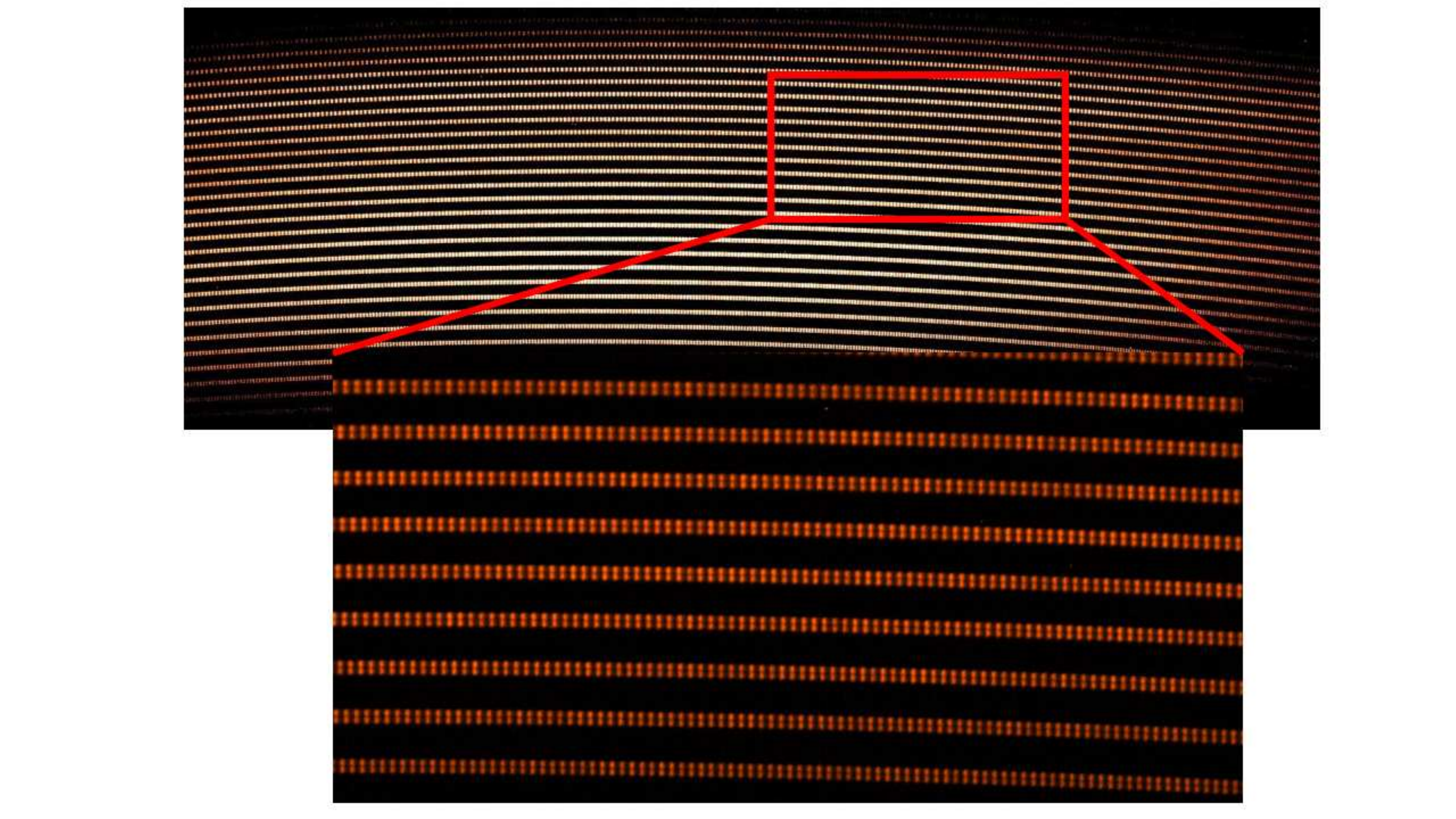}}
\caption{FP spectra observed at R=60000.The FP setup operates at a temperature of 19$^{\circ}$C $\pm$ 0.05$^{\circ}$C and under a presure of 0.025 mbar. Top panel: Full two dimensional spectra. Bottom panel: Zoomed-in version showing part of the spectra and the slicer image.\cite{ref:fpartifact}}
\label{fig:fullspectra}
\end{figure}

HESP is a high-resolution general purpose spectrograph installed on the 2m Himalayan Chandra Telescope (HCT), located at Indian Astronomical Observatory (IAO), Hanle, at an altitude of 4500~m above sea level. HESP covers a wavelength range of 350-1000~nm and has been designed to carry out a wide variety of scientific studies, including the ability to conduct RV studies of exoplanet host stars \cite{ref:sriram}. It provides two modes of operation, low-resolution mode offering a resolution of R=30000 and high-resolution mode, which uses an image slicer, giving a resolution of R=60000. A passively stabilized FP based wavelength calibration system has been installed on HESP spectrograph \cite{ref:fpao}. The FP spectra obtained in high-resolution mode (R=60,000), is shown in figure \ref{fig:fullspectra}. IRAF and a Python based pipeline, which does not incorporate tilt correction, is used for reduction of data obtained from the spectrograph.

\section{Methodology and algorithm}
\label{sec:method}

We have developed a routine that corrects the curvature in spectra and removes the tilt of the individual spectral lines. Image processing tools in Python, namely, Scikit-Image \cite{ref:scikitimage} and OpenCV \cite{ref:opencv} were used for this purpose. 

\subsection{Aperture Tracing}
\label{subsec:trace}
The first step in the process is to identify the position of all the apertures in the spectra. The position of all these apertures are saved as higher-order polynomials (2 or above), depending on which, the apertures are then extracted. This is called tracing of the apertures, and a continuous spectrum like that of a flat lamp is used for this purpose. Tracing of apertures was performed following the routines described in CERES \cite{ref:ceres}.

\subsection{Order extraction}
\label{subsec:extract}
\begin{figure}[htbp]
\centering
\vbox{\includegraphics[width=0.98\linewidth]{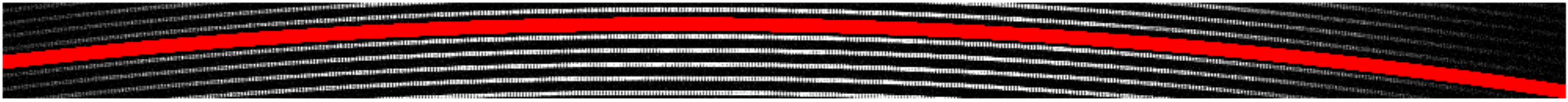}}
\caption{Aperture extraction mask, in red, over plotted on the spectra, for one particular aperture. There is no overlap of the mask with neighbouring orders.}
\label{fig:mask}
\end{figure}

After all the apertures were traced, the apertures are extracted in the next step. The aperture size ($apsize$) was chosen based on the extent of every aperture in the cross dispersion direction. While choosing the aperture size, the inter-order separation in the spectra is kept in mind in order to avoid any overlapping of the mask with consequent orders. Intensity values in the aperture were extracted by creating a mask using Bivariate Spline \cite{ref:zhou} and sampling at every 0.5 pixels. Mask extent was decided from $(y0 - apsize/2)$ to $(y0 + apsize/2)$ in step size of 0.5, where y0 is the traced polynomial. An example of the traced mask is shown in figure \ref{fig:mask}.

\subsection{Curvature removal}
\label{subsec:curvature}
\begin{figure}[htbp]
\centering
\vbox{\includegraphics[width=0.98\linewidth]{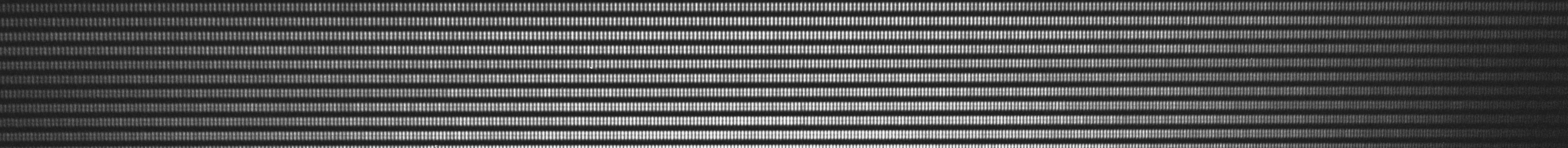}}
\caption{Section of the stacked apertures to show the curvature correction.}
\label{fig:str_spectra}
\end{figure}

Correction of the tilt in individual spectral lines should be preceded by curvature and global tilt removal, if any. For removing the curvature and straightening the new array, extracted points were stored in a separate array row-wise. The intensity data extracted from the first row was stored in the first row of the new array and so on. Stacking each straightened aperture in order of their extraction reproduced the entire spectra without the curvature as shown in figure \ref{fig:str_spectra}. 

\subsection{Tilt calculation and removal}
\label{subsec:tilt_remove}
The tilt is now computed in curvature removed results. Routine was tested on both FP and Th-Ar calibration frames (in X-Shooter and MIKE). Since the spectra have lines with distinguishable edges, we have used the Canny edge detection technique \cite{ref:canny} for determining the boundaries of each tilted line. The correction was performed individually on each aperture. The selection, tilt calculation and correction procedures are entirely automated. The algorithm for tilt correction is as follows:
\begin{itemize}
    \item Load the required aperture and take a central y-cut through the middle of the aperture.
    \item Smoothen the data obtained from y-cut using Gaussian smoothening. Detect the peak values in smoothened data and save the x-position corresponding to the peaks. This generates position information for all the lines present.
    \item Calculate the difference between adjacent peak positions. This gives the separation between two peaks. Take the median value of differences obtained. Construct a boundary array with $x peak position$ $\pm$ $median/2$ being edge positions. This is done to avoid overlapping of each FP line boundary when the final corrected array is reconstructed.
    \item Use the boundary array to construct a box and isolate the region of interest in a separate array. Calculate the y-extent of the separated box to include areas with FP data present and avoid background noise.
    \item Since Th-Ar spectra does not have equally spaced lines of uniform intensity, mean thresholding is performed, and a binary image is generated, example of which is shown in figure \ref{fig:bin_thresh}. The transitions from 0 to 1 and vice versa determine the boundary array values mentioned in the previous step.
    \begin{figure}[htbp]
    \centering
    \vbox{\includegraphics[width=0.98\linewidth]{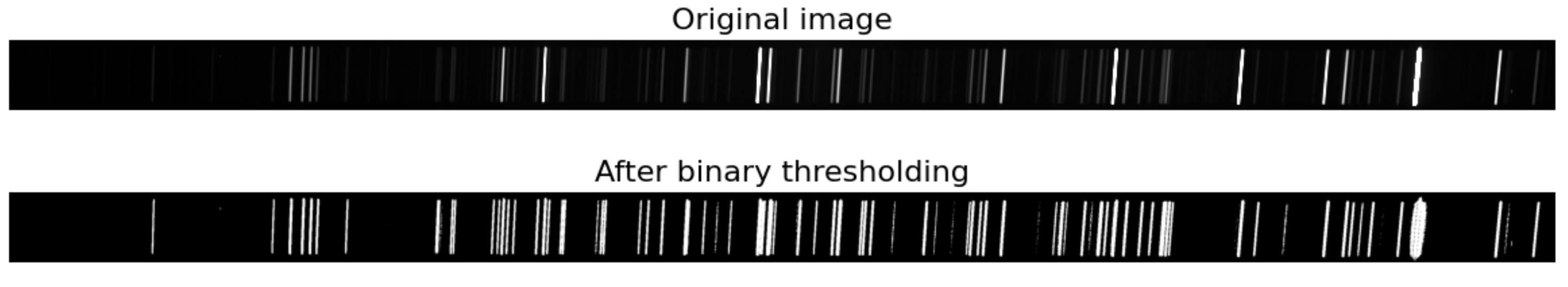}}
    \caption{Binary thresholding performed on one of the apertures of X-Shooter Th-Ar calibration frame. This is done to detect medium SNR lines. Top: Original extracted aperture. Bottom: Result of binary thresholding.}
    \label{fig:bin_thresh}
    \end{figure} 
    
    \item The centroid and tilt angle are determined using function region prop in Scikit processing package on each calibration line. The centroid value is taken as the reference position (x0,y0).
    \item Every pixel is shifted with respect to the reference position (x0, y0) determined above, using Eq. \ref{eq:shift} \cite{ref:dubs2015,ref:dubs2016}. Function warp in the scikit-image processing tool is used for shifting pixels according to the calculated $x_s$ value.
    
    \begin{equation}
    \label{eq:shift}
    x_s = x^\prime - (y^\prime-y_0) \times slope \, ,
    \end{equation}
    
    \item The result of the algorithm applied on one tilted line is shown in figure \ref{fig:tilt_correct}. Top panel shows the corrected image and bottom panel shows the effect of correction on the FWHM of the line. The corrected line is stitched into the reconstructed main array using the boundary values determined before. This is done to make sure that the FSR of the data is not altered.

\begin{figure}[htbp]
\centering
\subfloat[Tilt correction for single line]{\includegraphics[width = 0.7\linewidth]{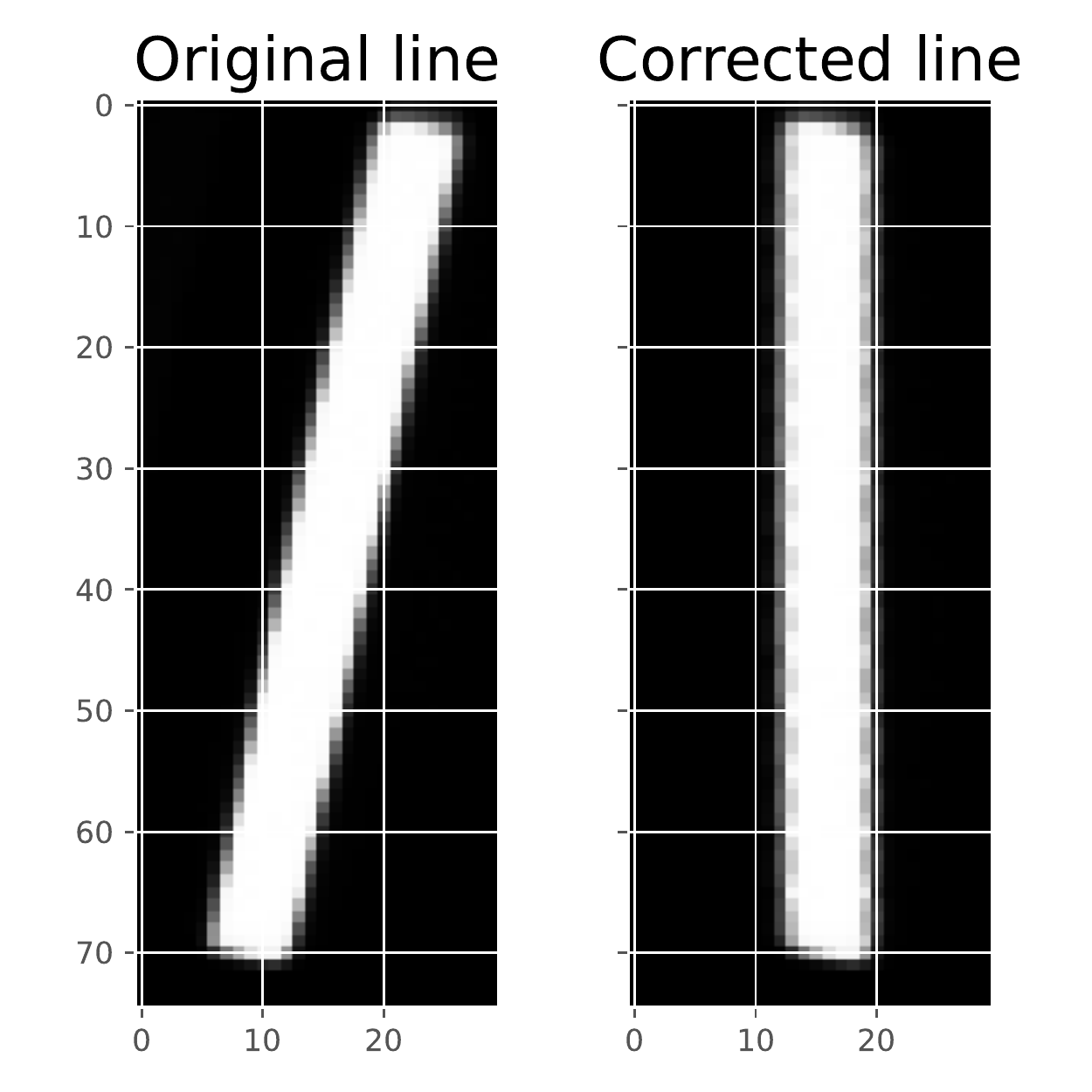}}\\
\subfloat[1D binned plot]{\includegraphics[width = 0.8\linewidth]{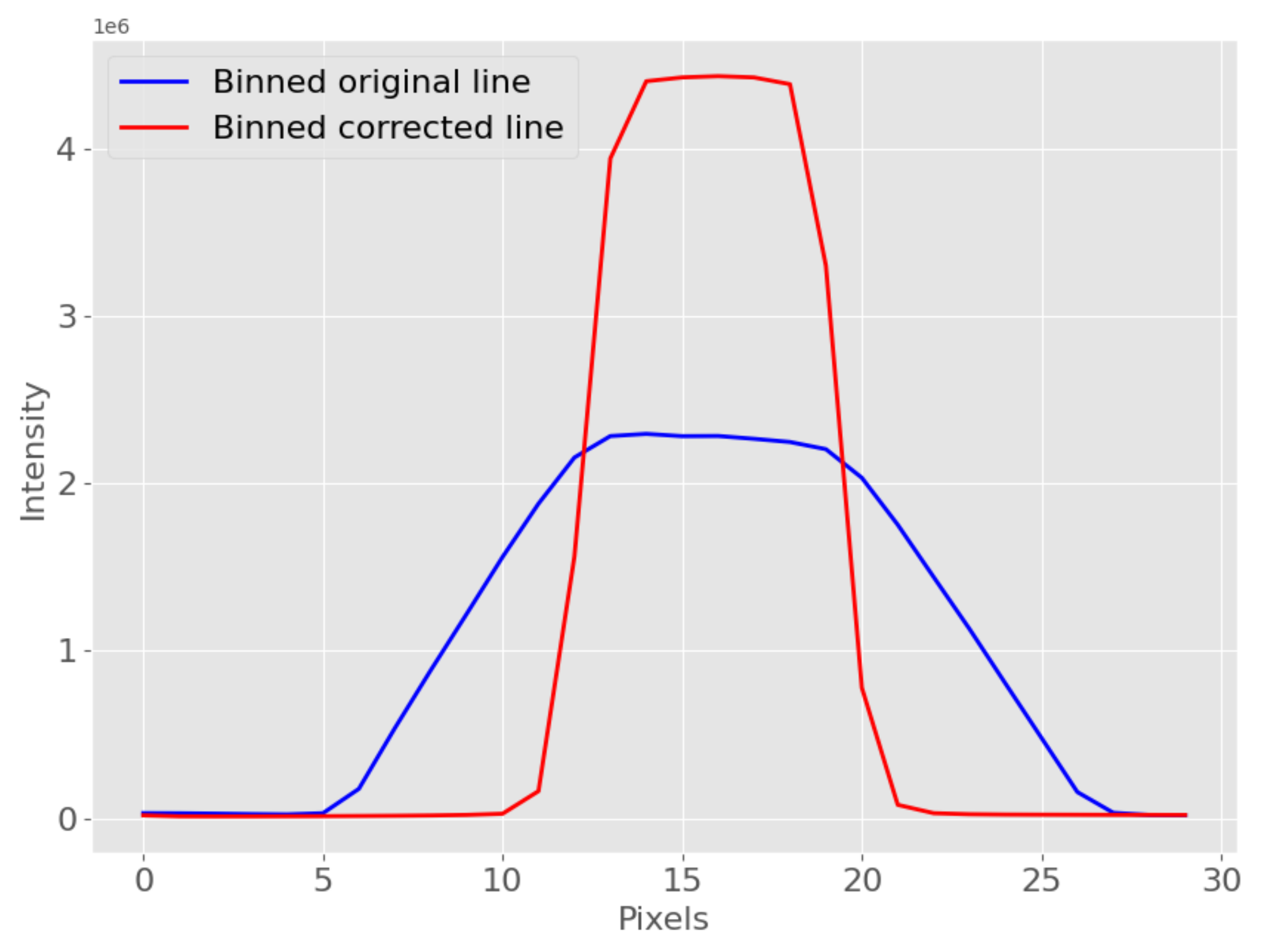}}
\caption{Representation of tilt correction. (a) Correction algorithm performed on one Th-Ar line to show how the tilt is removed. Original line shows the tilted line and corrected line is the result after correcting for tilt. (b) Binning done across y-axis to generate 1D plot for both the lines. It can be observed that the tilt correction drastically reduces the line-width from 12.44 before correction to 6.14 after correction.}
\label{fig:tilt_correct}
\end{figure}

\end{itemize}

\section{Results}
\label{sec:result}

The developed method has been tested on Fabry-Perot calibration spectra of HESP. For X-shooter and MIKE, Th-Ar calibration spectra were used.

\subsection{HESP}
\label{subsec:hespresult}
\begin{figure}[htbp]
\centering
\vbox{\includegraphics[width=0.98\linewidth]{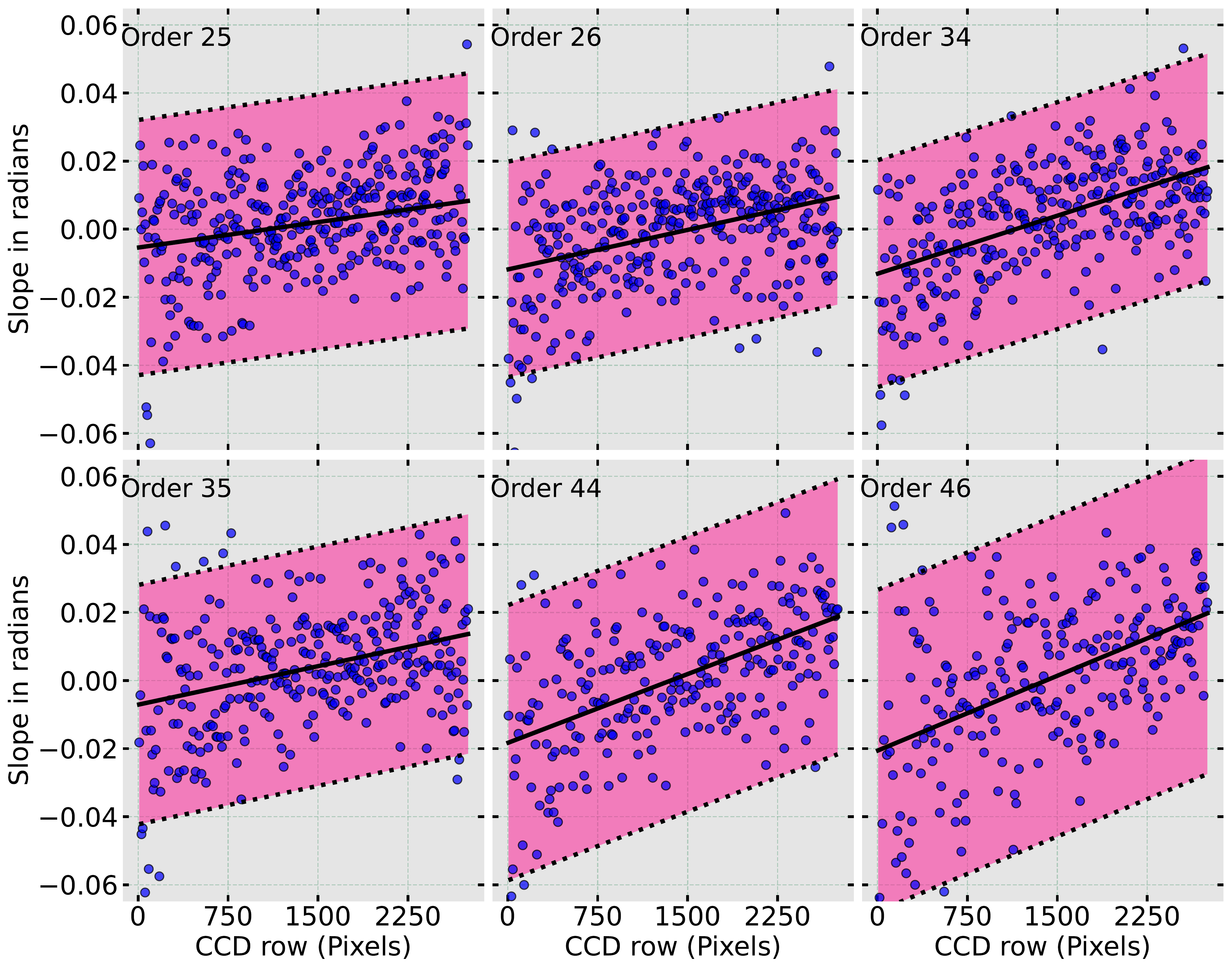}}
\caption{Tilt angle variation of FP lines across several spectrograph orders. The solid black line shows a linear fit to the data, and the dotted line shows a two sigma clipping boundary. Any point outside this boundary is rejected as an outlier during the fit. The scatter in the plot indicates the low tilt value in HESP spectra.}
\label{fig:hesp_slope}
\end{figure}

The FP calibration frame of HESP has been shown in figure \ref{fig:fullspectra}. Tilt calculation is preceded by curvature removal and global tilt correction described in the previous section. FP lines exist across 22 orders of the spectra, and in figure \ref{fig:hesp_slope} we show the representative plots for six orders: two top orders, two central orders and two bottom orders. The calculated tilt for FP lines is plotted along with a linear fit to the points. In HESP, we do not see a smooth variation in tilt angle after removal of curvature. As illustrated in figure \ref{fig:hesp_slope}, the weak trend in the data, however, is still visible from small but non-zero slopes of the best fit lines seen in different orders. The wavelength dependence of tilt is indicated by the slope of the linear fit to the tilt values, which increases from lower-order (\num{7.75e-6} for order 25) to higher order (\num{1.46e-5} for order 46). The tilt values are more scattered because the HESP design does not introduce a large tilt in the individual lines, causing the noise level in the spectra to come into the picture. For spectrographs in which it is known that the design introduces appreciable tilt, the scatter in the tilt angles are less, as will be seen in forthcoming sections.

\begin{figure}[htbp]
\centering
\vbox{\includegraphics[width = \linewidth]{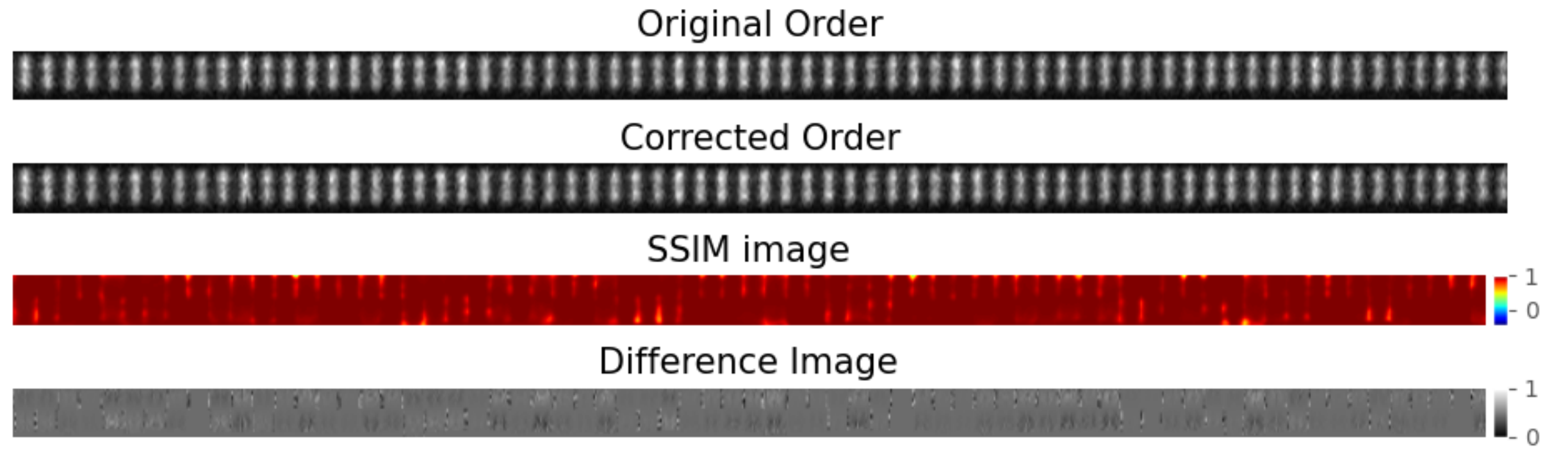}}
\caption{Zoomed in view of one tilt corrected FP spectral order, taken with HESP. Original order is the image before tilt correction. Corrected order is the image after correction. SSIM image shows the similarity image, 1 being the parts where the images are similar. The similarity index for the order shown is 0.987469. The difference image shows the normalized mathematical difference computed between two images, 0 being no difference regions.}
\label{fig:hesp_ssim}
\end{figure}

Figure \ref{fig:hesp_ssim} shows the uncorrected and corrected orders, plotted along with their similarity and difference image for one order. Structural Similarity Index (SSIM) is used to determine the similarity between two images, +1 indicating the most similar images and -1 indicating the images are very different \cite{ref:ssim}. SSIM image and difference image are computed between the original order and corrected order. 

\begin{figure}[htbp]
\centering
\vbox{\includegraphics[width = \linewidth]{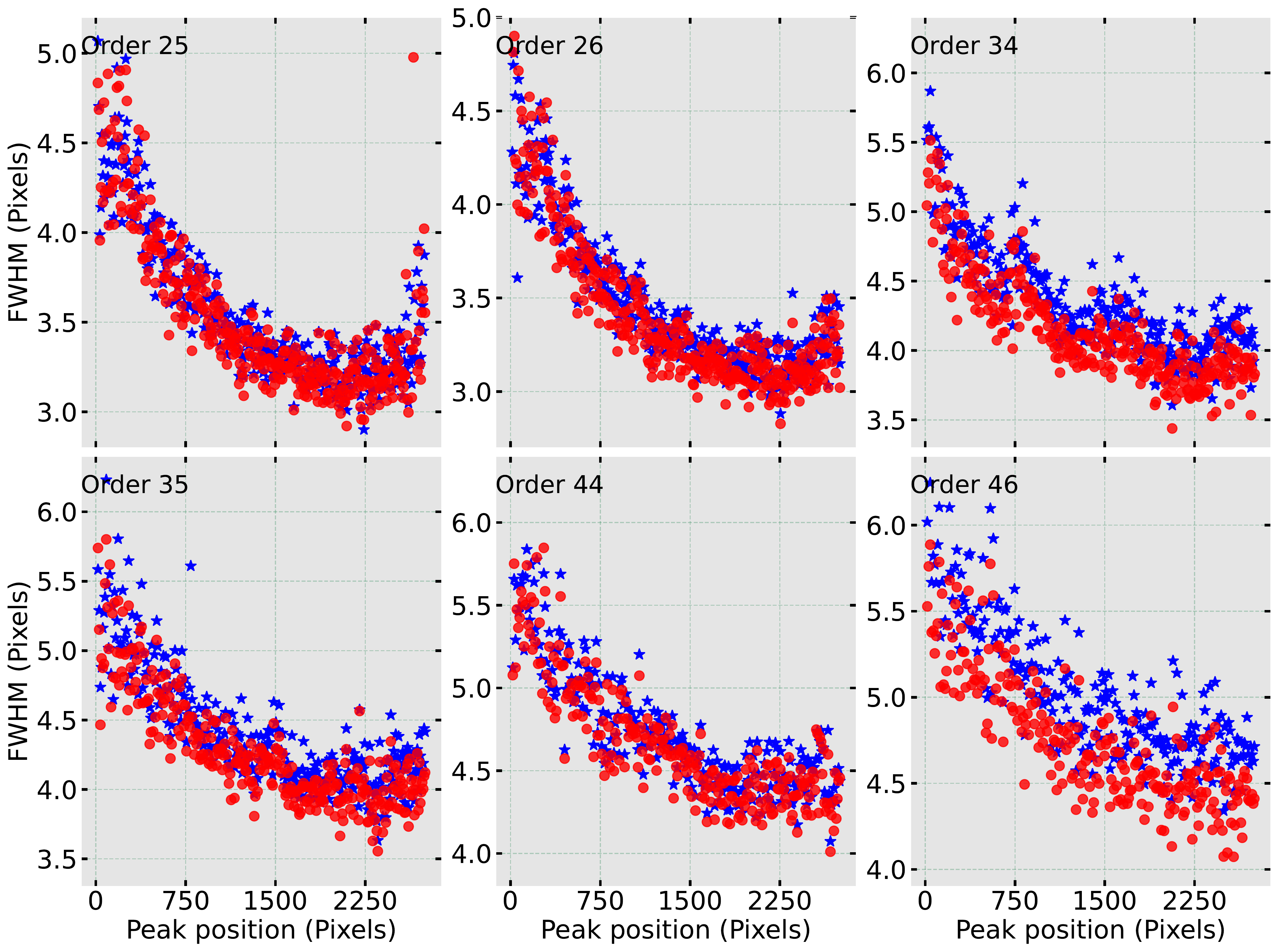}}
\caption{Comparison of FWHM of FP lines across an order before and after tilt correction. Values marked with blues stars show the FWHM of lines before correction. Values marked with red circles show the FWHM values of the same lines after correction.}
\label{fig:hesp_comp}
\end{figure}

Since HESP does not show any visible tilt, the FWHM of each FP line across the order is plotted in figure \ref{fig:hesp_comp} before and after the correction. The number of lines that show a decrease in FWHM is determined and the finesse is calculated. Not all the FP lines show a decrease in FWHM, especially the lines at edges, which can show an increase. This is because of low SNR at the edges due to the non-uniform illumination of the detector. Hence the correction is not performed effectively at these regions. The FWHM and finesse values for the six plotted orders are tabulated in table \ref{tab:fwhm}.

\begin{table}[htbp]
\centering
\caption{\bf Summarised result from tilt correction on HESP data.}
% \begin{tabular}{cccccc}
\begin{tabular}{|p{0.7cm} |p{1cm}| p{1.2cm}| p{0.7cm}| p{1cm} |p{1cm}|}
\hline
Order no. & Original mean finesse & Corrected mean finesse & Total peaks & Peaks with increased FWHM & Peaks with decreased FWHM\\
\hline
25 & 2.161 & 2.175 & 360 & 75 &	285\\
26 & 2.250 & 2.300 & 355 & 45 & 310\\
34 & 2.053 & 2.151 & 309 & 2 & 307\\
35 & 2.057 & 2.115 & 304 & 14 & 290\\
44 & 2.291 & 2.326 & 254 & 11 & 243\\
46 & 2.257 & 2.387 & 243 & 0 & 243\\

\hline
\end{tabular}
\label{tab:fwhm}
\end{table}

\subsection{X-shooter}
\label{subsec:xshresult}

\begin{figure}[htbp]
\centering
\subfloat[Order 9:SSIM=0.778193]{\includegraphics[width = 0.98\linewidth]{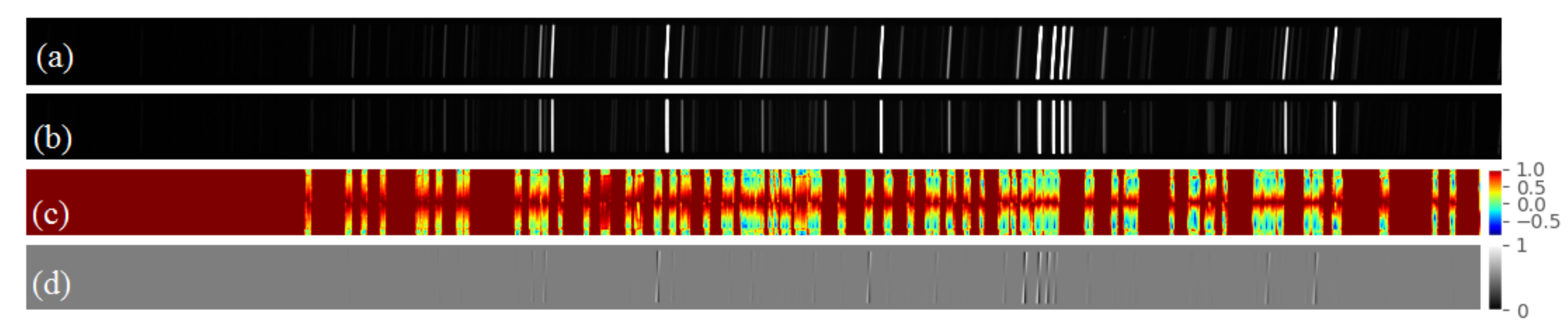}}\\
\subfloat[Order 10: SSIM=0.749655]{\includegraphics[width = 0.98\linewidth]{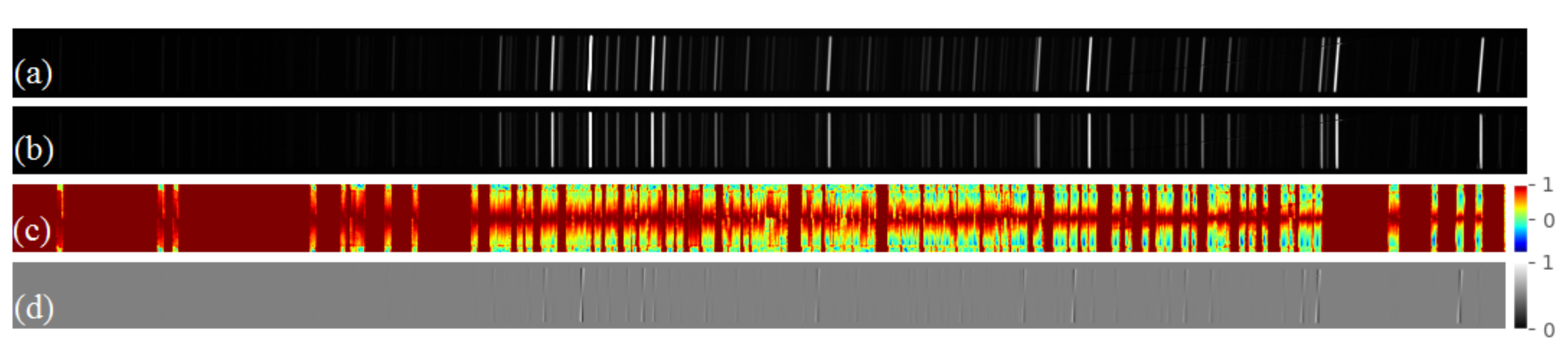}}
\caption{Tilt correction of Th-Ar spectra taken with X-Shooter. (a) Original spectrum before the tilt correction. (b) Th-Ar spectrum after tilt correction. (c) SSIM image showing the similarity image. The regions with highest similarity in the images corresponds to 1. (d) Difference image shows the normalized mathematical difference computed between two images, 0 being no difference regions. The similarity index is mentioned for each order.}
\label{fig:xsh_ssim}
\end{figure}

Th-Ar calibration frames have been used to test the tilt correction algorithm. Unlike FP, Th-Ar does not provide equispaced lines of uniform intensity, which makes line detection across the order tricky. The method applies well on high to medium SNR arc lines. Figure \ref{fig:xsh_ssim} shows the images of uncorrected and corrected orders, plotted along with their similarity image and difference image calculated between the original order and the corrected order. 

\begin{figure}[htbp]
\centering
\vbox{\includegraphics[width=\linewidth]{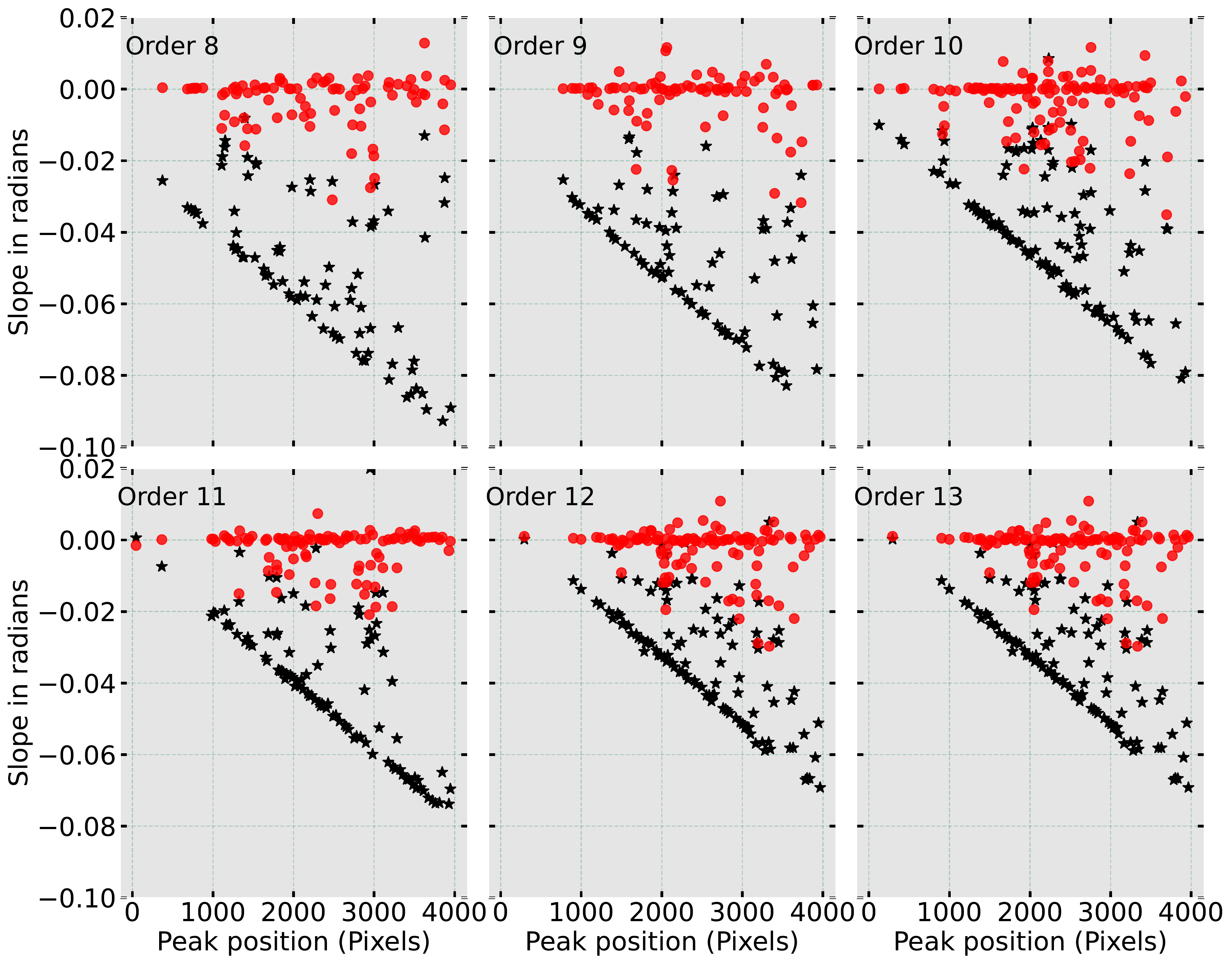}}
\caption{Slope values plotted for the uncorrected (black stars) Th-Ar lines and the corrected lines (red circles) for X-Shooter. The uncorrected slope values follow a clear trend. The scatter arises due to the attempt to calculate slopes of low SNR lines. The slope of the Th-Ar lines has been reduced after correction, the only deviation arising around the low SNR lines.}
\label{fig:xsh_slope}
\end{figure}

X-shooter introduces visible tilts in the spectra, with the tilt amplitude being $\sim$ 1.5-2 times the HESP tilt values. Hence the calculated tilt shows less scatter than in HESP. Figure \ref{fig:xsh_slope} shows the comparison between the slope values before and after correction. Most of the outliers in the data coincide with low SNR lines.

\subsection{MIKE}
\label{subsec:mikeresult}

\begin{figure}[htbp]
\centering
\subfloat[2" slit width Order 21:SSIM=0.933707]{\includegraphics[width = 0.98\linewidth]{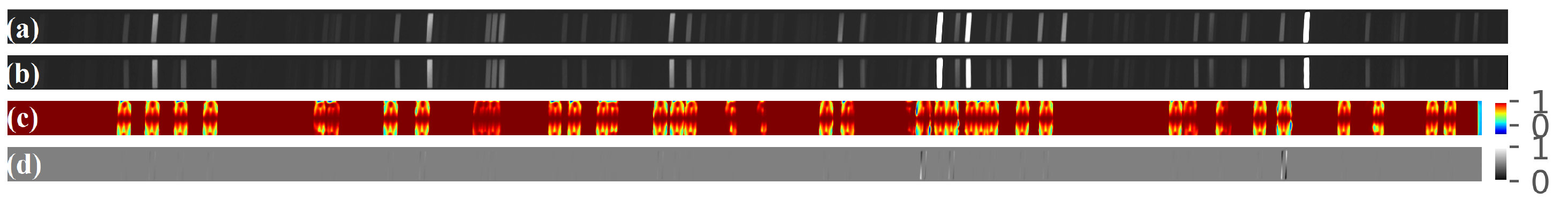}}\\
\subfloat[0.7" slit width Order 21: SSIM=0.946387]{\includegraphics[width = 0.98\linewidth]{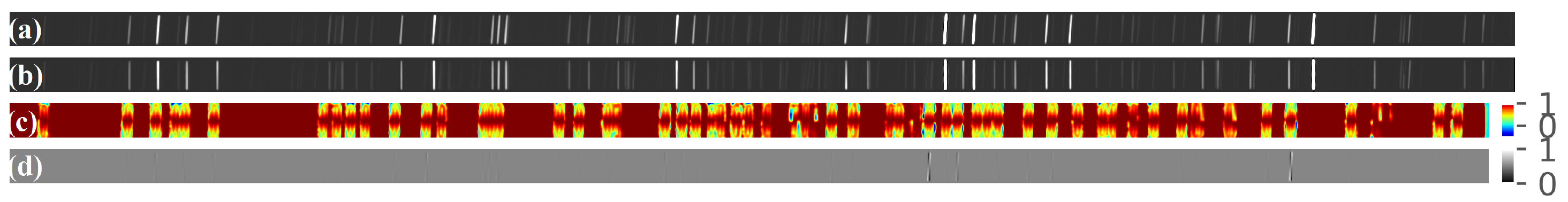}}\\
\subfloat[0.35" slit width Order 21: SSIM=0.944193]{\includegraphics[width = 0.98\linewidth]{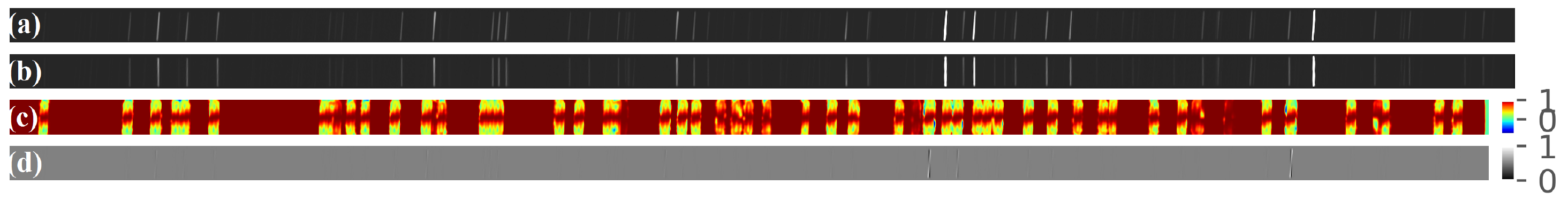}}
\caption{Tilt correction of Th-Ar spectra for different slit width setting of MIKE in red channel. (a) The image of order before tilt correction. (b) The image of same order after correction. (c) SSIM image shows the similarity image, 1 being the areas with maximum similarity. (d) Difference image shows the normalized mathematical difference computed between two images, 0 being no difference regions.}
\label{fig:imdiff}
\end{figure}

\begin{figure}[htbp]
\centering
\vbox{\includegraphics[width=\linewidth]{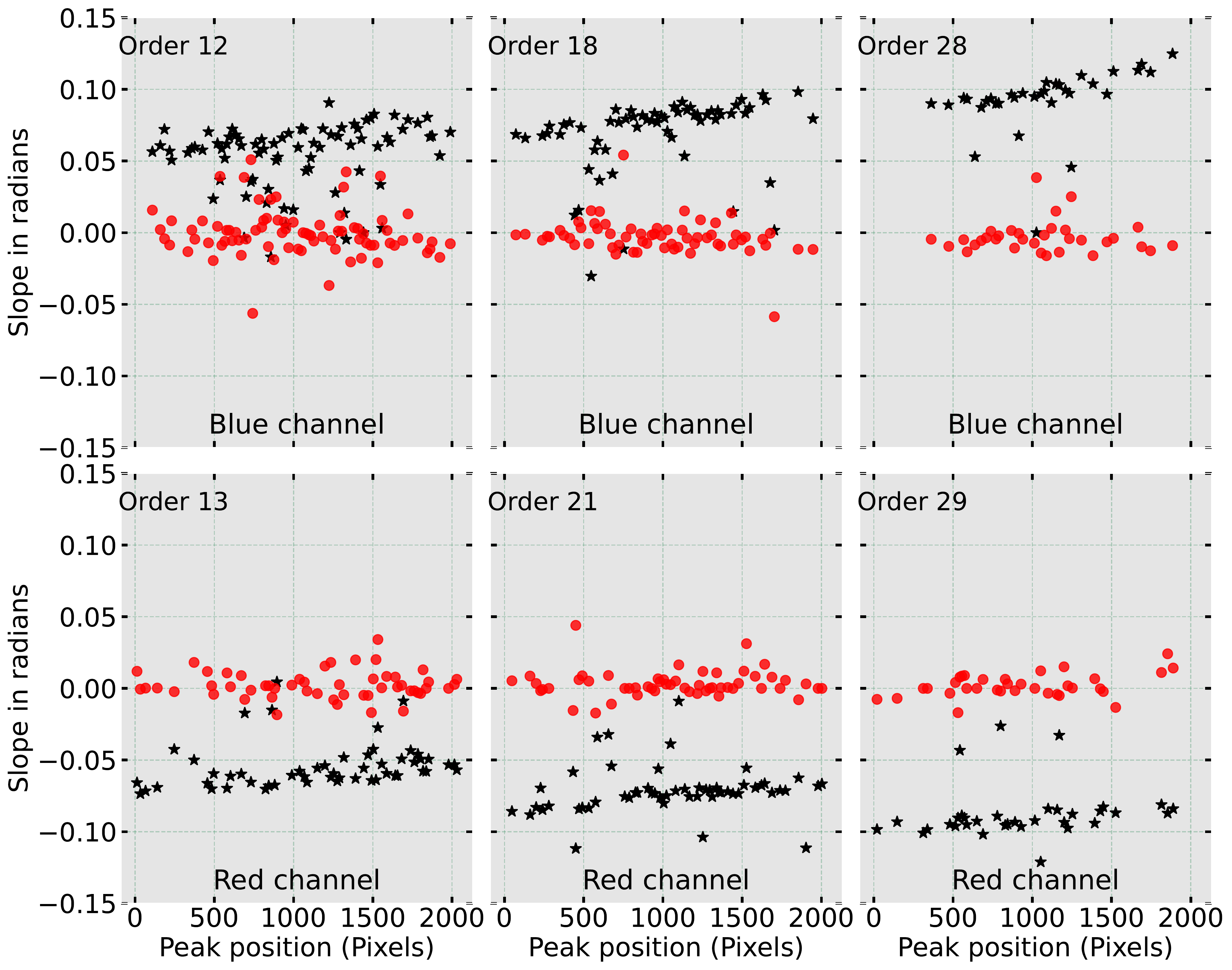}}
\caption{Slope values plotted for the uncorrected (black stars) Th-Ar lines and the corrected lines (red circles) for 2" slit setting of MIKE. The uncorrected slope values show a clear trend. The slope of the Th-Ar lines has been reduced after correction. The scatter in data is due to the attempt to calculate and correct for the low SNR lines.}
\label{fig:m2_slope}
\end{figure}

\begin{figure}[htbp]
\centering
\vbox{\includegraphics[width=\linewidth]{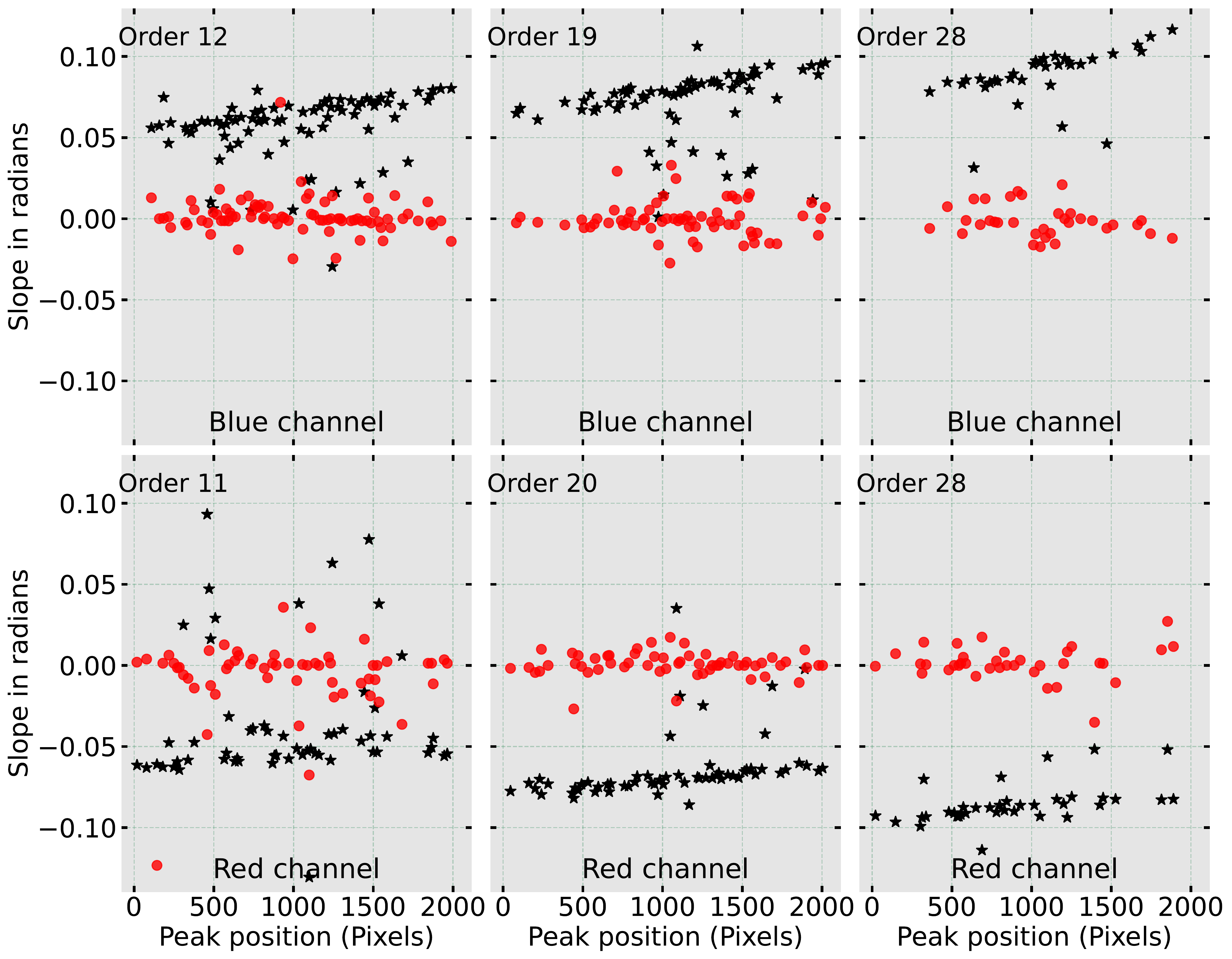}}
\caption{Slope values plotted for the uncorrected (black stars) Th-Ar lines and the corrected lines (red circles) for 0.7" slit setting of MIKE. The uncorrected slope values show a clear trend. The slope of the Th-Ar lines has been reduced after correction. The scatter in data is due to the attempt to calculate and correct for the low SNR lines.}
\label{fig:m7_slope}
\end{figure}

\begin{figure}[htbp]
\centering
\vbox{\includegraphics[width=\linewidth]{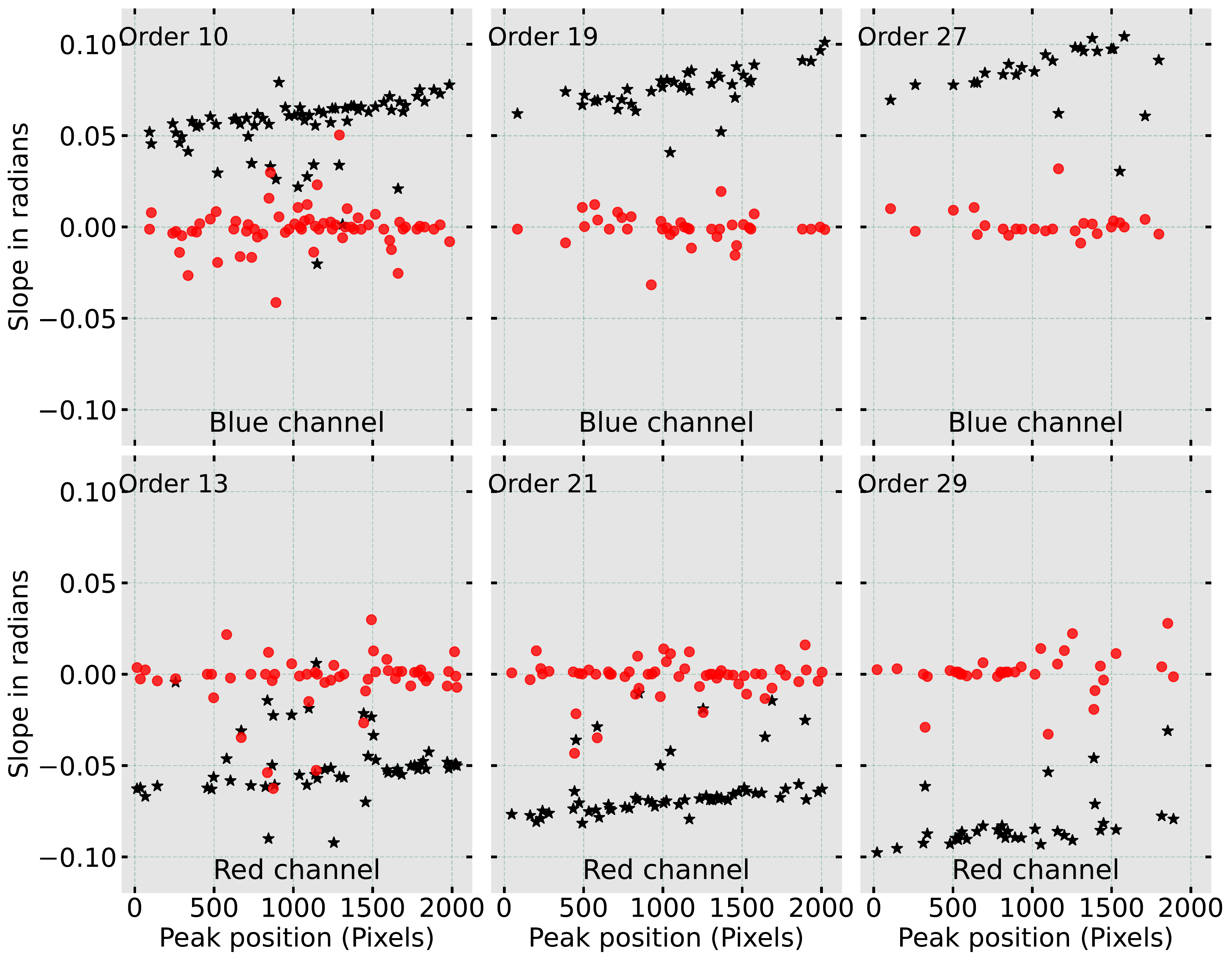}}
\caption{Slope values plotted for the uncorrected (black stars) Th-Ar lines and the corrected lines (red circles) for 0.35" slit setting of MIKE. The uncorrected slope values show a clear trend. The slope of the Th-Ar lines has been reduced after correction. The scatter in data is due to the attempt to calculate and correct for the low SNR lines.}
\label{fig:m35_slope}
\end{figure}

Since Th-Ar spectra do not share the same properties as FP spectra, only high to medium SNR lines are detected and corrected. We have performed the analysis on both blue and red channels of the spectrograph and data with slit settings of 2", 0.7" and 0.35" and will be presenting results of all the settings mentioned for the top, middle and bottom order. Figure \ref{fig:imdiff} shows the corrected and difference images for the mentioned slit settings of spectra in the red channel. The tilt in lines before and after correction is visible in the figure. Figure \ref{fig:m2_slope} shows the slope values before correction and after correction for spectra with 2" slit width. Figure \ref{fig:m7_slope} shows the corrected slope values for 0.7” slit data. The corrected  slope values for 0.35” slit setting is plotted in figure \ref{fig:m35_slope}. All the corrected slope plots are shown for both blue and red channels.

\begin{figure}[htbp]
\centering
\vbox{\includegraphics[width=\linewidth]{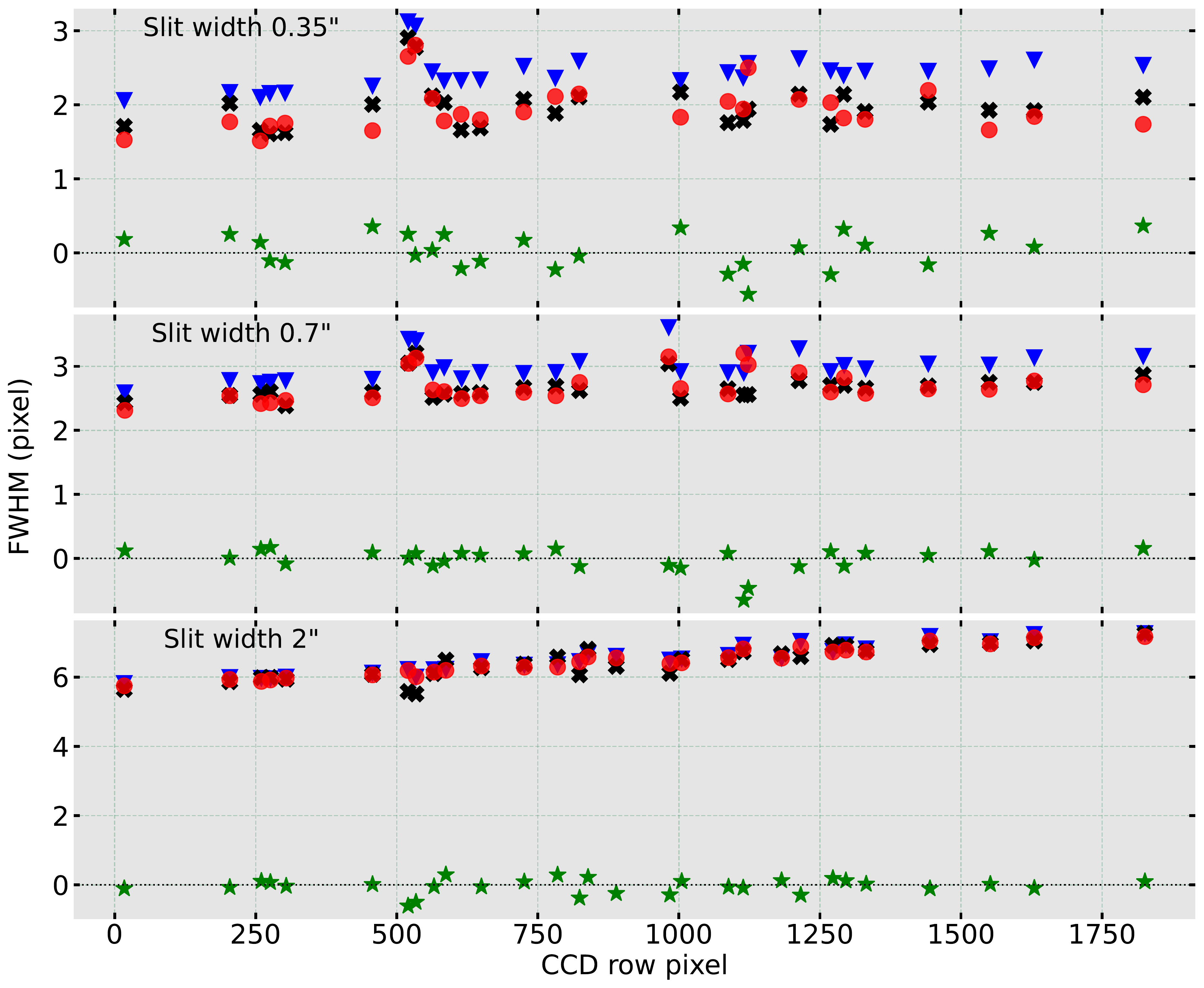}}
\caption{Performance of the developed algorithm with respect to MIKE pipeline. The blue triangles represent the FWHM of uncorrected tilted lines. The black crosses indicate FWHM of lines obtained after reduction with MIKE pipeline. The red circles indicate the FWHM of lines obtained after reduction with our algorithm. The green stars indicate the FWHM differences between the MIKE pipeline and our algorithm.}
\label{fig:pipe_comp}
\end{figure}

Figure \ref{fig:pipe_comp} shows comparison between the FWHM of extracted Th-Ar lines by MIKE pipeline ($lw1$) and by the discussed algorithm ($lw2$). The difference between $lw1$ and $lw2$ has also been plotted, a positive difference obtained when $lw1$ is greater than $lw2$. The values have been calculated for one order of Th-Ar spectrum obtained in red channel using slit widths of 0.35'', 0.7'' and 2''. For 0.35'' and 0.7'' slit widths, the pipeline gives 19\% $\pm$ 7\% and 11\% $\pm$ 4\% mean reduction in the FWHM of the Th-Ar lines respectively. For the same slit widths, the discussed algorithm gives 20\% $\pm$ 7\% and 10\% $\pm$ 4\% mean reduction in the FWHM respectively.

\section{Summary}
\label{sec:summary}
Analysis of the FP spectra acquired through HESP led us to the observation of discrepancies between the expected FP line FWHM (obtained with FTS) and the FWHM of the lines obtained with the spectrograph. This motivated us to examine the factors that can cause the increase in line-width. Performance of the FP degrades over the years due to deterioration of the optical coating. An additional reason for the increase in line-width was found to be the tilt in the FP lines, caused due to curvature in the orders. The HESP design does not introduce any significant tilt in the data. Hence it is not visually discernible in the spectra. Although the amount of the tilt is small, it is important to take care of this artifact in post-processing in order to maintain the performance of FP.

Tilted lines in echelle spectra are artifacts often introduced by the curvature in spectra or due to the design of the spectrograph itself. It is essential to eliminate this artifact in order to avoid the introduction of any errors in the data while post-processing. A possible way is to avoid binning the data and analyzing the flux at a single-pixel location. However, this technique will result in lesser flux values and hence less SNR in the data. Binning the 2D spectrum in the usual way is done vertically, i.e. along the slit direction. When a tilted line is binned vertically, it results in wrong flux values at each binned pixel as well as a decrease in spectral resolution due to broadening of the line, and hence wrong wavelength when the final calibration is performed. The tilt is not constant and varies as a function of wavelength across the spectra. This also affects the accuracy with which we can determine the centroid positions of spectral lines, which is crucial in the case of high precision RV studies. To avert the miscalculations arising from binning a tilted line vertically, either the binning could be performed along an oblique axis or a simpler way is to find out the obliquity of each spectral line and correct for it. We employ the latter method to remove the tilt by automating the detection of each spectral line, computing the line's slant, and then compensating for it, so that the standard post-processing techniques can be applied as usual.

We present in this paper a simple algorithm for curvature and tilt correction in high-resolution spectra. The algorithm is written in Python and uses image processing techniques for finding individual line slant values and correcting for them. We have demonstrated the algorithm on FP calibration spectra from HESP. A total of 6624 FP lines were detected across 22 orders, and a reduction in FWHM of 5417 lines was observed after using the tilt correction algorithm along with an overall improvement in the finesse value from 2.167 to 2.208. In order to ascertain the capability of the algorithm, we looked into spectrographs with highly tilted lines and performed the tilt corrections on the Th-Ar calibration spectra from X-shooter and MIKE. We noticed visual improvement in tilts of the lines after the application of the algorithm. We also calculated the tilt values of each line before and after correction and found a reduction in the absolute value of the slope after correction, thereby indicating the efficacy of the algorithm. The main limitation of the algorithm is the ability to deal with low SNR/faint lines (SNR<6) and blended
lines in Th-Ar spectra since it has been developed, keeping in mind the uniform line density provided by FP or LFC. We are looking into it as part of a future upgrade. Authors can be contacted to avail the code.

% \textbf{Currently, the main limitation of the algorithm is its inability to deal with very low SNR/faint lines (SNR<6) and blended lines in Th-Ar spectra which results in small but noticeable increase in FWHM. This is partly due to the fact that algorithm was primarily developed keeping in mind the uniform line density provided by FP or LFC. This issue can be addressed either by filtering out or setting an appropriate threshold on the SNR for correction. We plan to incorporate this feature in the future upgrade.}

\begin{backmatter}
\bmsection{Funding}
This project is supported by Science and Engineering Research Board (SERB), Department of Science and Technology (DST), India, under grant no. EMR/2014/000941. 

\bmsection{Acknowledgments}
The authors want to acknowledge Mr S. Sriram for the advice and technical support provided by him. The authors also thank Dr. Martin Dubs for his help and suggestions and Dr. Rebecca Bernstein for providing us with MIKE data. X-shooter study is based on data obtained from the ESO Science Archive Facility under request number 609319. 

\bmsection{Disclosures}
The authors declare no conflicts of interest.

\bmsection{Data Availability Statement}
Data underlying the results presented in this paper for X-shooter are publicly available on ESO Archive. The data for HESP and MIKE is not publicly available at this time but may be obtained from the authors upon reasonable request. 

\end{backmatter}

%%%%%%%%%%%%%%%%%%%%%%% References %%%%%%%%%%%%%%%%%%%%%%%%%
% \newpage

\bibliography{references}

\end{document}